\documentclass[conference]{IEEEtran}
\IEEEoverridecommandlockouts
\usepackage{cite}
\usepackage{amsmath,amssymb,amsfonts}
\usepackage{algorithmic}
\usepackage{graphicx}
\usepackage{textcomp}
\usepackage{xcolor}
\usepackage[ruled,vlined]{algorithm2e}

\def\BibTeX{{\rm B\kern-.05em{\sc i\kern-.025em b}\kern-.08em
    T\kern-.1667em\lower.7ex\hbox{E}\kern-.125emX}}
\begin{document}

\title{Temporal-Aware Iterative Speech Model for Dementia Detection\\}


\author{\IEEEauthorblockN{1\textsuperscript{st}Chukwuemeka Ugwu }
\IEEEauthorblockA{\textit{Systems Engineering Department} \\
\textit{ Stevens Institute of Technology}\\
 Hoboken, USA \\
cugwu@stevens.edu}
\and
\IEEEauthorblockN{2\textsuperscript{nd} Oluwafemi Oyeleke}
\IEEEauthorblockA{\textit{Systems Engineering Department} \\
\textit{ Stevens Institute of Technology}\\
 Hoboken, USA \\
ooyeleke@stevens.edu}}

\maketitle

\begin{abstract}

Deep learning systems often struggle with processing long sequences, where computational complexity can become a bottleneck. Current methods for automated dementia detection using speech frequently rely on static, time-agnostic features or aggregated linguistic content, lacking the flexibility to model the subtle, progressive deterioration inherent in speech production. These approaches often miss the dynamic temporal patterns that are critical early indicators of cognitive decline. In this paper, we introduce TAI-Speech, a Temporal Aware Iterative framework that dynamically models spontaneous speech for dementia detection. The flexibility of our method is demonstrated through two key innovations: 1) Optical Flow-inspired Iterative Refinement: By treating spectrograms as sequential frames, this component uses a convolutional GRU to capture the fine-grained, frame-to-frame evolution of acoustic features. 2) Cross-Attention Based Prosodic Alignment: This component dynamically aligns spectral features with prosodic patterns, such as pitch and pauses, to create a richer representation of speech production deficits linked to functional decline (IADL). TAI-Speech adaptively models the temporal evolution of each utterance, enhancing the detection of cognitive markers. Experimental results on the DementiaBank dataset show that TAI-Speech achieves a strong AUC of 0.839 and 80.6\% accuracy, outperforming text-based baselines without relying on ASR. Our work provides a more flexible and robust solution for automated cognitive assessment, operating directly on the dynamics of raw audio. 

\end{abstract}


\section{Introduction}

Dementia is a progressive neurodegenerative syndrome currently affecting an estimated 55 million people worldwide, with prevalence projected to rise sharply by 2050. It is marked by gradual decline in memory, language, and executive function, and Alzheimer’s disease remains the most common subtype \cite{OrtizPerez2023},\cite{Pan2025},\cite{Agbavor2022},\cite{Galanakis2025}. Early detection is critical for timely intervention and improved quality of life \cite{OrtizPerez2023},\cite{Gkoumas2024}. Among the most promising non-invasive biomarkers are speech and language changes, which often appear during preclinical stages \cite{Gkoumas2024},\cite{Agbavor2022},\cite{Pan2025},\cite{Li2025},\cite{Kannojia2025},\cite{Yeung2021}.

Speech deterioration is closely tied to functional decline measured by Instrumental Activities of Daily Living (IADLs) abilities such as financial management, medication adherence, and complex communication \cite{Fieo2014},\cite{Laurentiev2024},\cite{Fieo2018}. Extended IADL (x-IADL) scales correlate strongly with language function, processing speed, and visuospatial ability \cite{Fieo2014}. Despite extensive work analyzing speech or IADLs separately, current methods rarely model their temporal interdependence, even though language decline, commonly characterized as slowed speech, lexical retrieval failures, and reduced syntactic complexity, often precedes measurable IADL impairment \cite{Yeung2021},\cite{ChenLi2024}.

We hypothesize that gradual, fine-grained deterioration of speech is a precursor to IADL impairment and can be captured by an architecture inspired by optical-flow estimation. Both problems require tracking continuous temporal changes via correspondence analysis and iterative refinement \cite{Alfarano2024},\cite{Teed2021}. Analogous to how optical flow estimates motion between video frames, our approach models the temporal evolution of spectrogram frames, allowing precise characterization of pauses, pitch variability, and other subtle acoustic patterns.

We present TAI-Speech, a deep learning framework that treats speech as a dynamic sequence of spectrogram frames. Our model adapts the Recurrent All-Pairs Field Transform (RAFT) paradigm \cite{Alfarano2024},\cite{Teed2021},\cite{Sui2022} to audio analysis. A convolutional GRU serves as a recurrent update module that iteratively refines latent representations, while cross-attention aligns acoustic and prosodic cues. A Transformer encoder aggregates these temporally enriched features for utterance-level prediction. Unlike optical flow, we do not estimate motion vectors but leverage RAFT’s iterative refinement to construct a temporally aware embedding of the speech signal. Evaluated on the DementiaBank Pitt corpus, TAI-Speech outperforms strong linguistic baselines, demonstrating that temporally sensitive modeling substantially improves early dementia detection. While our conceptual framework highlights functional deterioration in IADLs, the current empirical evidence derives solely from speech-based classification; thus, the IADL connection remains theoretical.

\section{Related Work}
\subsection{Computational Approaches to Speech Based Dementia Detection}
\begin{figure}[htbp]
\centering
\includegraphics[height=5cm]{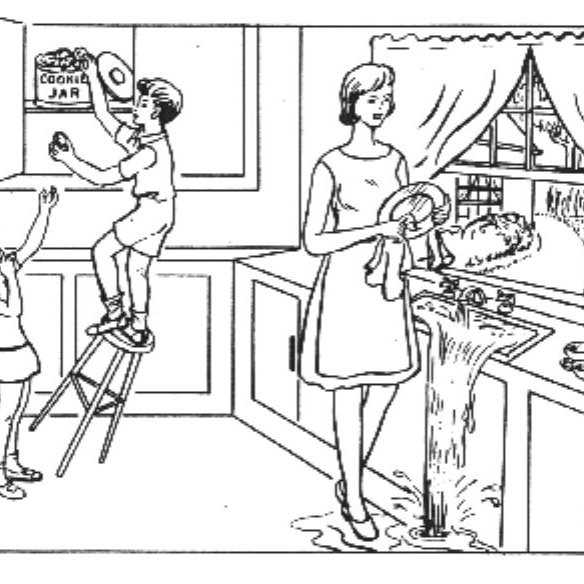}
\caption{Cookie Theft" picture}
\label{Prisma}
\end{figure}

Speech analysis has emerged as a non-invasive, cost-effective modality for early dementia diagnosis and monitoring \cite{OrtizPerez2023, Agbavor2022, Braun2024}. A cornerstone resource is the DementiaBank Pitt Corpus, which records subjects describing the “Cookie Theft’’ picture see Figure \ref{Prisma} to elicit lexical retrieval challenges and discourse impairments. Its derivatives, ADReSS and ADReSSo provide balanced demographics and higher acoustic quality, supporting tasks such as Alzheimer’s disease classification, MMSE regression, and cognitive-decline prediction, with ADReSSo emphasizing speech-only input and ASR-generated transcripts \cite{Luz2021}.

Feature extraction spans acoustic (log-Mel spectrograms, MFCCs, energy contours, pauses, hesitations) and linguistic (vocabulary richness, syntactic complexity, POS distributions, disfluency metrics) domains \cite{OrtizPerez2023,Braun2024,Ilias2023,Woszczyk2024}. Deep models dominate: CNNs for audio, RNN/LSTM and Transformer variants (e.g., BERT, RoBERTa, DeiT, GPT-3) for both acoustic and text representations \cite{Pan2025,Braun2024,Meilan2014,Koenig2018, Gong2021}. Self-supervised models such as wav2vec 2.0 capture rich acoustic embeddings with strong downstream performance \cite{Pan2025,Braun2024}.

Multimodal fusion strategies  integrate modalities through early feature concatenation, late decision-level aggregation, and cross-attention mechanisms that dynamically weight each modality. Although ASR errors can introduce noise, transcripts with relatively high Word Error Rates (WER) often perform on par with or better than manual transcriptions for dementia classification, suggesting that salient cognitive cues persist in noisy outputs \cite{Pan2025,Shon2023}.

\subsection{Correlating Speech with Functional Decline}
Loss of independence in Instrumental Activities of Daily Living (IADLs) is a defining clinical marker of dementia \cite{Fieo2014,LiepeltScarfone2013}. Modern, technology-mediated IADLs such as online financial tasks or text messaging offer even greater sensitivity for early Alzheimer’s detection \cite{Benge2024}. Numerous studies report strong links between speech abilities and functional status: language deficits often precede measurable IADL impairment  \cite{Gkoumas2024,Yeung2021}.

Automated speech and language analysis captures these associations objectively. Word-finding difficulty correlates with increased pause frequency and specific acoustic signatures (e.g., MFCC patterns), while incoherence and perseveration manifest as degraded discourse structure and repeated utterances measurable via cosine similarity. Reduced lexical diversity, simplified grammar, and malformed verb phrases signal syntactic and semantic breakdown \cite{Yeung2021}. Beyond structured tasks, NLP methods applied to unstructured clinical narratives in EHRs extract indicators of IADL/ADL impairment, enabling scalable integration of functional status into research and clinical decision support \cite{Laurentiev2024,Penfold2022}.

\subsection{Iterative Refinement in Optical Flow}

Optical flow estimation has evolved through iterative refinement, from early variational methods that balanced data fidelity and smoothness \cite{Alfarano2024,Teed2021}, to deep learning models that directly predict flow between frames \cite{Teed2021}. 
RAFT represents a major advancement by maintaining high-resolution flow fields and leveraging multi-scale correlation volumes with a convolutional GRU for iterative updates, outperforming traditional coarse-to-fine approaches \cite{Teed2021,Sui2022}.
Architectures like IRR build on FlowNetS and PWC-Net, using shared weights for refinement, while post-processing techniques such as bilateral filtering and CPF enhance flow quality by preserving edges and reducing noise \cite{Alfarano2024,Teed2021}. 
Models like LiteFlowNet further refine cost volumes through cascaded inference, and optimized loss functions like RCELoss enable efficient training with fewer parameters\cite{Alfarano2024,Hui2018}.These iterative principles, high-resolution correlation, recurrent updates, and context preservation inform broader temporal modeling strategies and motivate cross-domain applications beyond computer vision \cite{Alfarano2024, Messmer2025}.
\subsection{Theoretical Frame for Temporal Analysis}

Speech is a fundamentally temporal information modality, where the state at a given moment is intrinsically linked to its context. 
Temporal aspects in audio, such as hesitation and pauses, speaking rate, and word duration, serve as significant indicators of cognitive decline \cite{Xu2023}. Dementia recordings exhibit prolonged utterances and characteristic pause patterns, with manual transcripts often marking these events explicitly \cite{OrtizPerez2023},\cite{Pan2025},\cite{Braun2024}. Acoustic representations such as log-Mel spectrograms, MFCCs, and eGeMAPS capture short-term spectral and physiological voice dynamics \cite{OrtizPerez2023},\cite{Gong2021},\cite{Corvitto2024}, \cite{Luz2021}.

Longitudinal analysis tracks language change across sessions via embedding similarity and related metrics \cite{Gkoumas2024},\cite{Braun2024}. Linguistic deficits, empty speech, circumlocution, repetition, poor grammar are temporal manifestations of cognitive decline \cite{ChenLi2024}. Extra-linguistic cues such as keystroke pauses in written text further complement audio evidence \cite{Gkoumas2024}.

Context-aware large language models can exploit preceding audio or text to predict next-sentence semantics or topic flow, enhancing downstream temporal reasoning \cite{Shon2023},\cite{Bai2024}. Related techniques in audio-visual segmentation similarly rely on temporal consistency, where optical flow provides low-level motion signals for tasks like emotion recognition and lip-reading \cite{Alfarano2024},\cite{Torabi2014}. Temporal Enhancement Modules (TEM) extend these ideas by exchanging learnable context tokens across frames to strengthen inter-frame coherence \cite{Geng2025}.

\section{Methodology}

\subsection{Task and Dataset}
We evaluate our approach on spontaneous picture description, a standardized neurocognitive paradigm used to probe semantic memory and episodic retrieval (\cite{Mueller2018}). Participants describe the Cookie Theft line drawing (\cite{Lanzi2023}), producing naturalistic speech that reveals lexical retrieval difficulty, hesitations, and discourse-level impairments.

Experiments use the DementiaBank Pitt Corpus, the largest publicly available speech dataset for cognitive‐impairment assessment. We focus on the clinically validated subsets comprising 222 recordings from 89 healthy controls (HC) and 255 recordings from 168 participants with Alzheimer’s disease (AD), for a total of 477 audio samples. All recordings are accompanied by diagnostic annotations and are sampled at 16 kHz.

\subsection{Model Architecture}

\begin{figure}[htbp]
\includegraphics[height=8.0cm]{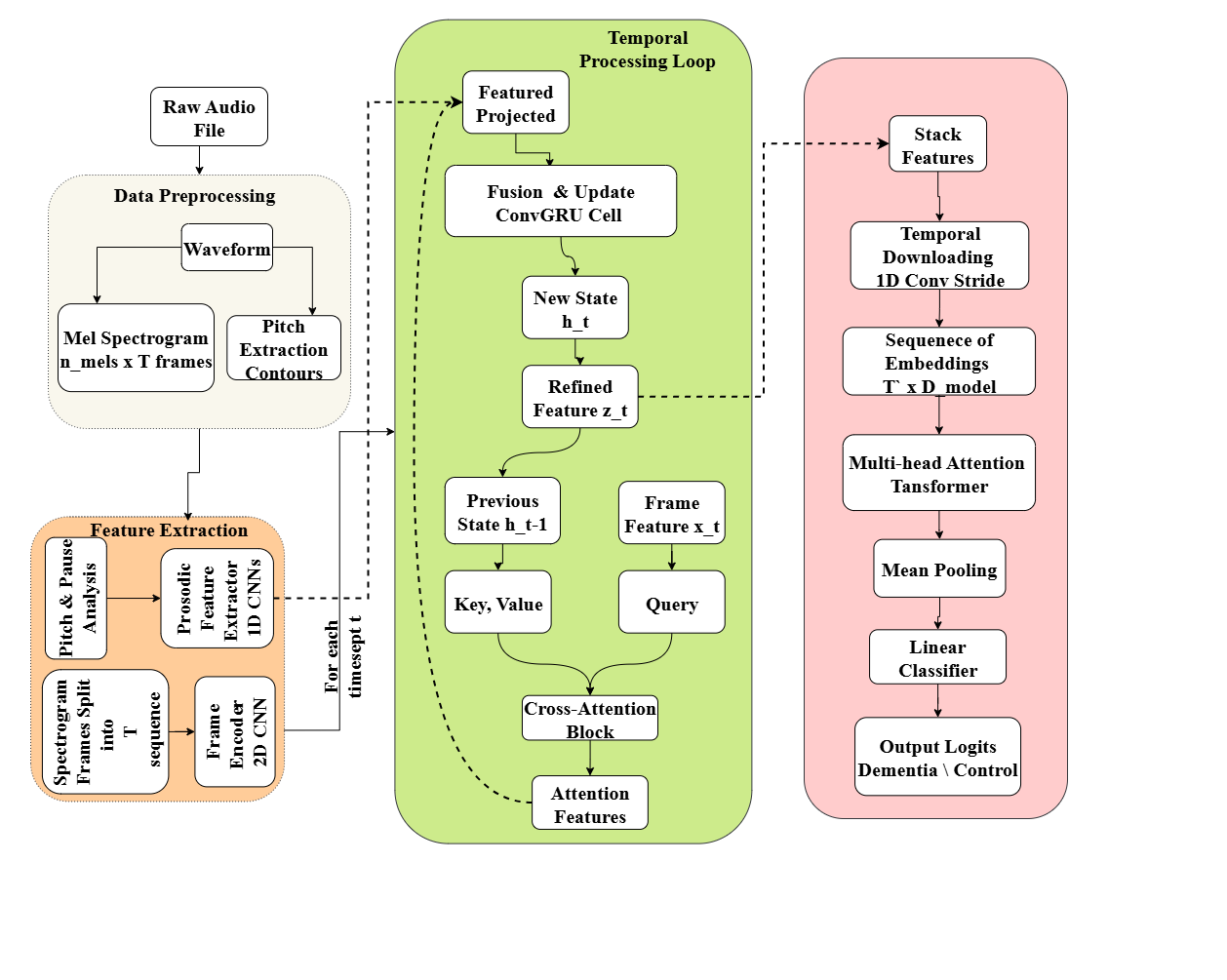}
\caption{Architecture Model }
\label{model_arch}
\end{figure}

Our goal is to capture temporal markers of functional decline, particularly those linked to Instrumental Activities of Daily Living directly from raw speech. \textbf{TAI-Speech} integrates prosodic encodings, convolutional spectral processing, iterative temporal refinement, and sequence-level aggregation (Figure \ref{model_arch}). The model is trained end-to-end with a joint objective combining cross-entropy classification and a temporal smoothness regularizer to enforce stability across successive frames.

\subsection{Feature Encodings}
Raw speech $x(t)$ is first resampled and transformed using the short-time Fourier transform (STFT). A log-Mel spectrogram is computed:
\begin{equation}
S(m,n) = \log \big( \sum_{k} |X(k,n)|^2 H_m(k) \big),
\end{equation}
where $X(k,n)$ is the STFT coefficient at frequency $k$ and frame $n$, and $H_m$ is the $m$-th Mel filter. 

Prosodic correlates relevant to IADL are explicitly extracted: (i) normalized pitch track $\tilde{p}(n)$, and (ii) pause probability $q(n)$ estimated from voice activity detection. These auxiliary encodings are fused into a joint representation:
\begin{equation}
z(n) = \phi\big(W_f [\tilde{p}(n), q(n)] + b_f\big),
\end{equation}
where $W_f$ and $b_f$ are trainable parameters and $\phi(\cdot)$ is a non-linear activation.

\subsection{Temporal Refinement Modules}

\subsubsection{Cross-Attention Contextualization}
The spectral encoder produces local embeddings $h^{(l)}$. To integrate prosodic factors, a cross-modal attention module computes:
\begin{equation}
\text{Attn}(Q,K,V) = \text{softmax}\!\Big(\frac{QK^\top}{\sqrt{d}}\Big)V,
\end{equation}
where queries $Q$ are derived from spectro-temporal features, while keys and values come from $z(n)$. This aligns acoustic features with temporal dynamics of pitch and pause, providing contextualized embeddings.

\subsubsection{Iterative Update Block}
To refine temporal representations, we employ a multi-scale ConvGRU. At each time step $t$, the hidden state $H_t$ is updated by:
\begin{align}
r_t &= \sigma(W_r * x_t + U_r * H_{t-1}), \\
u_t &= \sigma(W_u * x_t + U_u * H_{t-1}), \\
\tilde{H}_t &= \tanh(W * x_t + U * (r_t \odot H_{t-1})), \\
H_t &= u_t \odot H_{t-1} + (1 - u_t) \odot \tilde{H}_t,
\end{align}
where $*$ denotes convolution and $\odot$ elementwise multiplication. This iterative block progressively corrects and stabilizes features across multiple scales, reflecting temporal organization in speech.

\subsection{Sequence Aggregation and Classification}
Downsampled embeddings $\{h_1, \ldots, h_T\}$ are passed into a Transformer encoder augmented with a classification token $u_{\text{cls}}$. The self-attention mechanism models higher-order dependencies:
\begin{equation}
U = \text{Transformer}([u_{\text{cls}}, h_1, \ldots, h_T]).
\end{equation}
The final classification is computed as:
\begin{equation}
\hat{y} = \text{softmax}(W_c u'_{\text{cls}} + b_c),
\end{equation}
where $u'_{\text{cls}}$ is the contextualized embedding.

The training objective combines cross-entropy loss with a temporal consistency regularizer:
\begin{equation}
\mathcal{L} = \lambda_{\text{cls}} \mathcal{L}_{\text{CE}}(\hat{y}, y) + \lambda_{\text{temp}} \frac{1}{T-1}\sum_{t=2}^T \|h_t - h_{t-1}\|_2^2,
\end{equation}
encouraging stability in temporal encodings while preserving discriminative capacity.

\section{Experimental Setup}

\subsection{Evaluation Protocol}

In order to guarantee a rigorous and unbiased evaluation of the proposed approach, we adopt a stratified five-fold cross-validation (5-fold CV) protocol. This strategy preserves the original class distribution within each fold, a critical consideration when working with imbalanced clinical datasets. The primary evaluation metric is the Area Under the Curve (AUC), which provides a robust measure of discriminative capability between dementia and healthy control groups. In addition, we report secondary performance indicators, namely accuracy, precision, recall, and F1-score, thereby offering a comprehensive assessment across multiple dimensions of classification performance.

\subsection{Baseline System}

For comparative analysis, we adopted a strong literature-based baseline and evaluated against several Transformer-style architectures that constitute current state-of-the-art approaches for cognitive-impairment detection.
\subsection{Proposed System}

\SetKwComment{Comment}{\# }{}
\begin{algorithm}[t]
\caption{TAI-Speech: Temporal–Acoustic–IADL Speech Classification}
\label{alg:taispeech}
\KwIn{Raw waveform $x(t)$,\; ground-truth label $y$}
\KwOut{Predicted probability $\hat{y}$}
\Comment{Preprocessing}
Resample $x(t)$ and compute log-Mel spectrogram
$S(m,n)=\log \sum_k |X(k,n)|^2 H_m(k)$ \\
Extract normalized pitch $\tilde{p}(n)$ and pause probability $q(n)$ \\
Fuse prosodic vector $z(n)=\phi(W_f[\tilde{p}(n),q(n)]+b_f)$

\Comment{Spectral Encoding} 
$h(l)\leftarrow$ Hierarchical convolutional encoder on $S(m,n)$

\Comment{Cross-Attention Contextualization} 
$h'(l)\leftarrow \mathrm{Attn}(Q,K,V)$ with
$Q=h(l)$,\; $K,V=z(n)$

\Comment{Iterative Temporal Refinement} 
\For{$t=1$ \KwTo $T$}{
$r_t=\sigma(W_r*x_t+U_r*H_{t-1})$ \\
$u_t=\sigma(W_u*x_t+U_u*H_{t-1})$ \\
$\tilde{H}_t=\tanh(W*x_t+U*(r_t\odot H_{t-1}))$ \\
$H_t = u_t\odot H_{t-1} + (1-u_t)\odot \tilde{H}_t$
}

\Comment{Sequence Aggregation and Classification} 
$U \leftarrow \text{Transformer}([u_{\text{cls}}, h'_1,\dots,h'_T])$ \\
$\hat{y} \leftarrow \operatorname{softmax}(W_c u'_{\text{cls}} + b_c)$

\Comment{Training Loss} 
$\mathcal{L} = \lambda_{\text{cls}}\mathcal{L}_{\text{CE}}(\hat{y}, y)
+ \lambda_{\text{temp}} \frac{1}{T-1}\sum_{t=2}^{T}\|h_t - h_{t-1}\|_2^2$

\Return{$\hat{y}$}
\end{algorithm}

\paragraph{Algorithm}
Algorithm 1 presents the overall procedure of our proposed method. The TAI-Speech framework refines acoustic representations of spontaneous speech to detect dementia-related functional decline. 
The procedure can be summarized in three stages:

\begin{itemize}
  \item \textbf{Acoustic Feature Encoding:} Raw audio $x(t)$ is converted into log-Mel spectrogram frames $S(m,n)$. 
  A hierarchical convolutional encoder extracts local spectral representations as initial feature maps.
  
  \item \textbf{Iterative Temporal Refinement:} Hidden states $H_t$ are updated with a multi-scale ConvGRU to capture long-range temporal context. 
  The prosodic characteristics, the normalized pitch $\tilde{p}(n)$ and the probability of pause $q(n)$, are fused using a cross-modal attention layer for richer temporal contextualization.
  
  \item \textbf{Sequence Aggregation and Classification:} Refined embeddings are downsampled and passed through a Transformer encoder with a learnable classification token $u_{\text{cls}}$. 
  A final linear layer with softmax outputs the dementia vs.\ control prediction. 
  Training employs a cross-entropy loss plus a temporal-smoothness regularizer to encourage frame-to-frame consistency.
\end{itemize}

\subsection{Training Details}

To ensure comparability across systems, all models are trained under a unified hyperparameter configuration. Training uses the AdamW optimizer with a batch size of four, a maximum of 200 epochs, and an initial learning rate of  of $1\times10^{-5}$. Early stopping halts training if the validation AUC fails to improve for 10 consecutive epochs, reducing the risk of overfitting to the limited clinical dataset. Cross-entropy loss serves as the objective function, and a WeightedRandomSampler balances class representation within each batch. All experiments are executed on a shared NVIDIA RTX A4000 GPU with 32 GB of memory to support efficient large-model training and stable batch processing.

\section{Results and Discussion}

This section presents the performance of our proposed architecture, contextualizes the findings by comparing them against established baseline models, and discusses the broader implications of our results, with a specific focus on how the model's design relates to the detection of functional decline.

\subsection{Quantitative Performance Analysis}

The proposed model was rigorously evaluated using a 5-fold cross-validation protocol. The primary metric for assessing the model's ability to discriminate between dementia and healthy control classes was the Area Under the Curve (AUC), with Accuracy (ACC), Recall (REC), and F1-score also reported for a comprehensive analysis.

As summarized in Table~\ref{tab:result}, our proposed architecture achieved a high level of discriminative performance, yielding an  test AUC of 0.839. The model obtained an accuracy of 80.55\%, recall of 0.890, and an F1-score of 0.813. 

\begin{table}[h]
    \centering
    \caption{The Result (\%) of our Model }
    \label{tab:result}
    \begin{tabular}{@{}cllll}
    \hline
    System&AUC &ACC & REC & F1-score  \\ \hline 
    Our Model & 83.9 & 0.81 &0.890& 0.813 \\
    \hline
    \end{tabular}
    
\end{table}
For evaluation, we benchmarked our system against previously reported baseline models, as well as Transformer-based acoustic approaches that analyze transcribed speech. 
\begin{table}[h]
\centering
\caption{Performance comparison between the proposed model}
\label{tab:performance}

\setlength{\tabcolsep}{4pt} 
\begin{tabular}{@{}cllllll@{}}
\hline
 \textbf{System}& Modality &\textbf{AUC (\%)} &\textbf{Acc (\%)}  & \textbf{Recall} & \textbf{F1-score} \\
\hline
 \cite{Pan2019} &Lingustic      & --&70.83 & 0.71 & 0.70 \\
 \cite{Pan2025}  &Multimodal      &-- & \textbf{82.56} & 0.83 & \textbf{0.83}  \\
 \cite{Braun2024} &Multimodal       & 77.2 & -- & -- & -- \\
Ours& Acoustic-Temporal & 83.9& 80.55 & \textbf{0.89 }& 0.83   \\
\hline
\end{tabular}
\end{table}
 
The TAI-Speech architecture achieved an AUC of 0.839, an accuracy of 80.55\%, a recall of 89.0\%, and an F1-score of 0.813. These results represent a significant improvement over purely linguistic baselines. 
When benchmarked against state-of-the-art systems in table \ref{tab:performance}, our model demonstrates good performance across all evaluation metrics. The 8\% improvement in AUC over \cite{Braun2024}.'s pause-infused text model (77.2\%) and competitive performance against \cite{Pan2025} attention-based multimodal system (82.56\% accuracy) underscore the efficacy of our temporatsively on acoustic signals without requiring error-prone ASR transcription or linguistic feature extraction.

 While systems incorporating ASR features achieve the highest AUC and accuracy, our purely acoustic model obtains the highest recall. It is notable that TAI-Speech achieves this level of performance without relying on a linguistic pipeline. This suggests that the temporal dynamics encoded within the acoustic signal contain sufficient information for effective dementia classification. This single-modality approach may offer advantages in robustness and simplicity, as it avoids potential cascading errors from ASR systems, which can struggle with the atypical speech patterns often present in clinical populations. The results indicate that direct modeling of speech dynamics is a viable and powerful alternative to multimodal approaches that require transcription.

\subsection{Discussion}

The performance of TAI-Speech can be attributed to its architectural design, which adapts principles from optical flow to model speech as a continuous, evolving signal. This approach facilitates the capture of micro-temporal variations in speech that are often associated with cognitive decline. The iterative refinement mechanism, inspired by the RAFT architecture, allows the model to progressively build a representation of speech dynamics across multiple time scales.

The model's convolutional GRU-based update module may be effective at capturing hesitation patterns, variable speech rates, and prosodic irregularities characteristic of speech produced by individuals with dementia. In contrast to approaches that analyze the final linguistic product, our model is designed to capture patterns in the speech production process itself. While semantic errors and vocabulary limitations are important markers captured by linguistic models, they represent the endpoint of a cognitive process. Our method, by modeling the temporal trajectory of speech, aims to detect subtle perturbations that may precede overt linguistic deterioration.

Although direct IADL measurements were not incorporated into this study, the established link between speech production deficits and functional decline provides an interpretive context for our results. The model's sensitivity to temporal speech features aligns with known correlations between communication difficulties and IADL impairment. For example, word-finding difficulties, which manifest acoustically as increased pause frequency and duration, can affect functional tasks that require verbal communication. The cross-attention mechanism aligns spectral embeddings with prosodic dynamics like pitch and pauses, which may encode information about the cognitive effort involved in speech planning and execution. The degradation of these processes, as captured by our model, may serve as an indicator of the executive dysfunction that can lead to IADL impairment.
\subsubsection*{Limitations and Future Directions}

Despite promising results, this study has several limitations. The findings are based on a constrained dataset from a single linguistic and cultural context, which may limit their generalizability. The cross-sectional nature of the data also precludes any assessment of the model's sensitivity to longitudinal disease progression. Furthermore, the absence of direct IADL measurements restricts the empirical validation of our model's relevance to functional decline. The model's performance on mild cognitive impairment (MCI) also remains an open question for future investigation.

Future work should aim to validate these findings on larger, more diverse, and longitudinal corpora. Incorporating patient IADL scores as an explicit modeling target could provide a more direct method for detecting functional decline. Exploring multimodal fusion, which would combine the temporal acoustic features from TAI-Speech with semantic embeddings from large language models, may also lead to improved robustness and performance. Finally, longitudinal studies are necessary to determine if changes in the model's output correlate with cognitive trajectories over time, potentially enabling the use of personalized baselines for early detection.

\section{Conclusion}

In this work, we introduced TAI-Speech, a novel architecture for dementia detection that models the temporal dynamics of speech. By adapting the principle of iterative refinement from the field of optical flow, our model analyzes the evolution of acoustic-prosodic features over the course of an utterance. On the DementiaBank Pitt corpus, TAI-Speech achieved an AUC of 0.839, demonstrating performance that is competitive with state-of-the-art multimodal systems without requiring linguistic transcription. This result validates temporal modeling as an effective approach for this task. While the link between the captured speech dynamics and functional impairment is theoretically grounded, future work is required to empirically validate this connection using datasets that include clinical IADL assessments.

\bibliographystyle{ieeetr} 
\bibliography{ref}

\end{document}